\documentclass[11pt]{article}
\usepackage{moriond,epsfig}

\def\be{\begin{equation}}
\def\ee{\end{equation}}
\def\bea{\begin{eqnarray}}
\def\eea{\end{eqnarray}}

\begin{document}
%%%%%%%% HAS BEEN SUPPRESSED \vspace*{4cm}
\title{KONDO RESONANCE IN A NANOTUBE QUANTUM DOT COUPLED TO A NORMAL AND A SUPERCONDUCTING LEAD}

\author{M.R. GR\"ABER, T. NUSSBAUMER, W. BELZIG, T. KONTOS and C. SCH\"ONENBERGER}

\address{Institut f\"ur Physik, Universit\"at Basel, Klingelbergstrasse 82, 4056 Basel, Switzerland}

\maketitle\abstracts{We report on electrical transport
measurements through a carbon nanotube quantum dot coupled to a
normal and a superconducting lead. The ratio of Kondo temperature
and superconducting gap $T_{K}/\Delta$ is identified to govern the
transport properties of the system. In the case of $T_{K}<\Delta$
the conductance resonance splits into two resonances at $\pm
\Delta$. For the opposite scenario $T_{K}>\Delta$ the conductance
resonance persists, however the conductance is not enhanced
compared to the normal state due to a relative asymmetry of the
lead-dot couplings. Within this limit the data is in agreement
with a simple model of a resonant SN-interface.}

Over the last years there has been growing interest in studying
electronic correlations and their interplay in mesoscopic systems.
Prime examples for such correlations are magnetism and
superconductivity. From an experimental point of view carbon
nanotubes offer a well-suited platform to investigate such
phenomena and test theoretical predictions. Recent advances in
nano-fabrication allow the preparation of intermediate-transparent
superconducting contacts on carbon nanotube quantum dots allowing
the observation of both superconducting and (magnetic) Kondo
correlations. The scenario of a quantum dot (QD) in between two
superconducting leads was investigated experimentally by Buitelaar
{\it et al.} identifying the ratio of Kondo temperature $T_{K}$
and superconducting order parameter $\Delta$ to govern the
interplay of both correlations \cite{buit2,buit3} in agreement
with theoretical works \cite{choi,zaikin}. In this article we
consider a slightly different setup with a nanotube quantum dot
coupled to one superconducting and one normal lead (S-QD-N). This
geometry has been studied extensively on a theoretical basis and
various predictions such as $\Delta$-sidepeaks, excess Kondo
resonances, enhancement or suppression of the conductance have
been made \cite{cuevasSN,clerkSN,excess,fazioSN}. However, in the
normal state the many-particle-phenomenon Kondo effect reduces
effectively to a non-interacting resonance for temperatures well
below $T_{K}$. This gives rise to the question whether this
picture is still valid in an S-QD-N geometry. We will thus in the
following compare our experimental data with the model of a
non-interacting resonance in between a normal and a superconductor
as given by Beenakker for the linear response regime and by Khlus
{\it et al.} for a finite bias voltage \cite{beenakker,khlus}.

\begin{figure}[t]
\begin{center}
\includegraphics [width=0.8\textwidth]{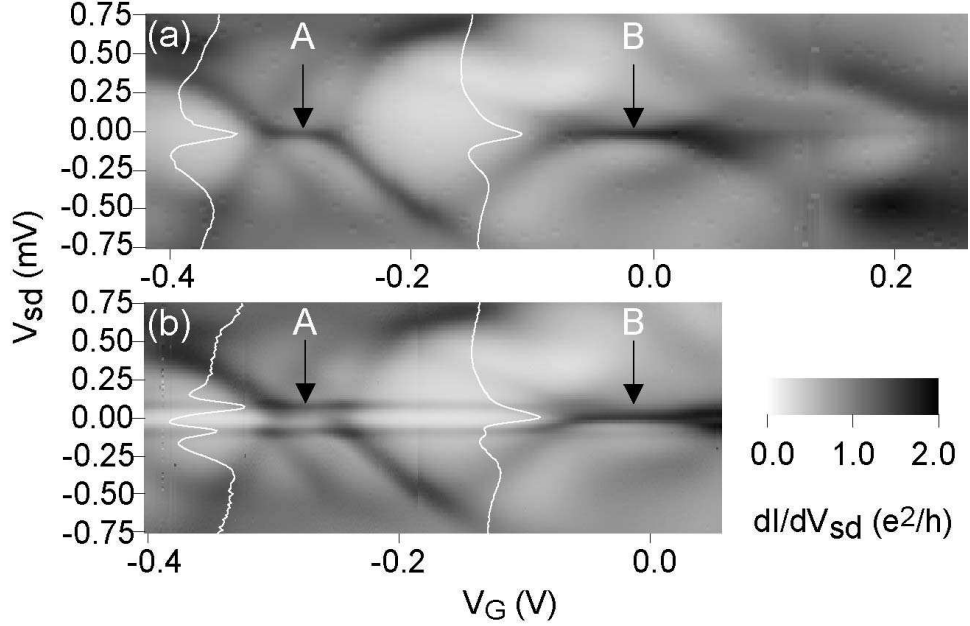}
\caption{(a) Greyscale representation of the normal state
conductance at 90 mK and B=25 mT (dark = more conductive). The
white curve on the left (right) shows the differential conductance
versus the applied source-drain voltage at the position of the
left (right) arrow. The two Kondo ridges are labelled ``A'' and
``B''. (b) Greyscale representation of the conductance in the
superconducting state at 90 mK and B=0 mT.} \label{figure1}
\end{center}
\end{figure}

\begin{figure}
\begin{center}
\includegraphics [width=0.8\textwidth]{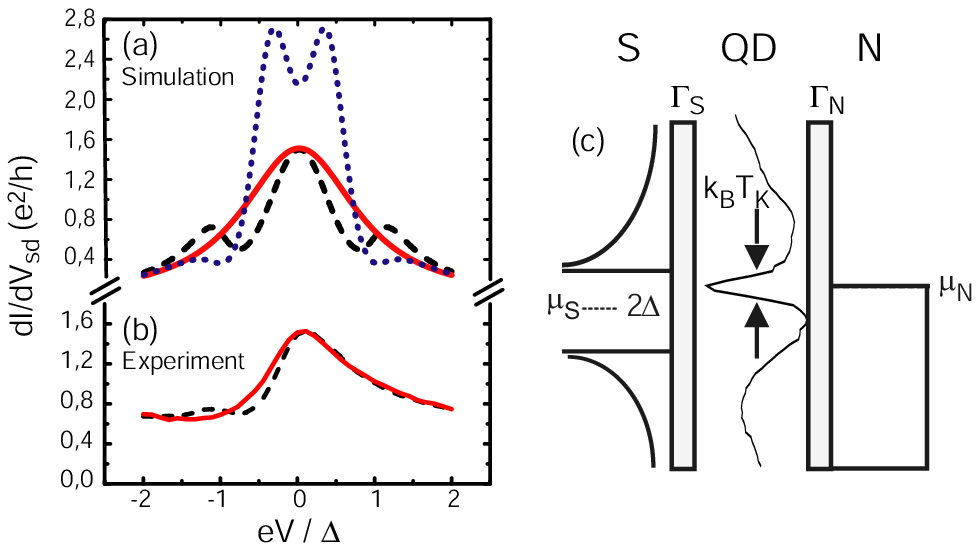}
\caption{(a) Calculated differential conductance versus
source-drain voltage at T=0.1 $\Delta$. The solid curve represents
the normal state with $\Gamma_{L/R}=0.60$ and
$\Gamma_{R/L}=0.37*0.6=0.22$. The dashed (dotted) curve
corresponds to the superconducting state and $\Gamma_{S(N)}=0.22$
and $\Gamma_{N(S)}=0.6$. (b) Measured differential conductance at
T=90 mK in the normal (solid line) and superconducting (dashed
line) state at the center of Kondo ridge ``B''. (c) Simplified
schematics of a quantum dot coupled to a normal and a
superconducting lead in the Kondo regime. } \label{figure2}
\end{center}
\end{figure}

We report on electrical transport measurements through a
multi-walled carbon nanotube (MWNT) contacted by short-spaced (300
nm) metal electrodes, thus acting as a quantum dot at low
temperatures. The superconducting contact is formed by a 45 nm Au/
160 nm Al proximity bilayer, the normal one by a 45 nm Au layer.
Further experimental details are described elsewhere
\cite{MattNT}. By applying a magnetic field of 25 mT being bigger
than the Aluminum critical field but small in terms of the Zeeman
shift of the nanotube levels the superconducting electrode is
driven into the normal state and the system can be characterized
in an N-QD-N configuration. Figure \ref{figure1}(a) shows a
greyscale representation of the differential conductance through
our device in the normal state at 90 mK. Clear traces of Coulomb
blockade diamonds for an even number of electrons on the dot and
high-conductive Kondo ridges around zero source-drain voltage in
between the diamonds are evident. The Kondo effect occurs when the
number of electrons on the dot is odd thus acting as a localized
magnetic moment with spin S=1/2. Below $T_{K}$ the spins in the
leads try to screen this localized moment by co-tunnelling
processes allowed only on a short timescale within the Heisenberg
uncertainty principle. This results in a resonance of the linear
differential conductance around zero source-drain voltage. The
Kondo temperature $T_{K}$ can be determined from the temperature
dependence of the linear differential conductance given by
$G(T)=G_{0}/(1+(2^{1/s}-1)(T/T_{K})^2)^s$ where s=0.22 for a spin
1/2 system \cite{Goldhaber-Gordon}. A best fit to the data yields
0.3 K and 1.3 K for the ridges labelled ``A'' and ``B'',
respectively. The maximum conductances in the unitary limit
$T<<T_{K}$ turn out to be $G_{0,A}=1.54 \;e^{2}/h$ and
$G_{0,B}=1.57 \;e^{2}/h$.

When one of the two electrodes enters the superconducting state
the Kondo effect becomes modified. It was shown for a S-QD-S
geometry that the Kondo effect is suppressed by superconductivity
when $\Delta>T_{K}$. In the present setup of an S-QD-N geometry
one faces an asymmetric situation with the Kondo processes between
normal lead and dot remaining unaffected by the onset of
superconductivity. However, this is not the case for the processes
between the superconducting electrode and the dot. In figure
\ref{figure2}(c) the electronic spectrum of the QD and the leads
is depicted. Depending on the ratio $\Delta/T_{K}$ there are two
possible scenarios for the superconductor-dot coupling. For
$T_{K}>\Delta$ the Kondo effect is expected to persist since
quasiparticle states in the superconducting lead can participate
in the Kondo spin flip processes. A natural question to ask now is
whether the presence of the superconducting electrode results in
an enhancement of the conductance through the quantum dot. Like
for any high transmission SN contact one would expect an enhanced
conductance of up to $4 \:e^{2}/h$ being the maximum conductance
of a perfectly-transmitting channel. If, however, $T_{K}<\Delta$
states around the superconductor chemical potential are missing
and the Kondo coupling between the dot and the superconducting
lead will be strongly suppressed. Yet a resonance with
renormalized $T_{K}<T_{K}*$ persists in the dot local density of
states which is pinned to the normal lead chemical potential. Due
to the effective lowering of the dot-superconductor coupling in
this scenario Andreev reflections become less probable and the
conductance resonance at zero source-drain voltage vanishes. On
the other hand one expects additional features at
$V_{sd}=\pm\Delta/e$, i.e. when the remaining Kondo resonance and
the BCS singularities in the density of states are lined up.

Figure \ref{figure1}(b) shows the differential conductance through
our device in the superconducting state at T=90 mK and B=0 mT. The
magnitude of the superconducting gap can be deduced from the
horizontal feature at $eV_{sd}=\Delta\approx 0.09$ meV. The Kondo
ridges ``A'' and ``B'' thus represent the two cases described
above with $T_{K,A}/\Delta\approx 0.3$ and $T_{K,B}/\Delta\approx
1.3$. As expected for ridge ``A'' the resonance in the linear
differential conductance does not persist but highly conductive
features at the position of the superconducting gap are apparent.
In contrast to the latter case the conductance resonance at zero
bias persists in the superconducting state for ridge ``B''.
However, besides a small feature at $V_{sd}=-\Delta/e$ the shape
and the absolute value remain almost unchanged. When fitting the
temperature dependence as described above one obtains a maximum
conductance of $G_{0,B}=1.55 \:e^{2}/h$. For the temperature
dependence of the measured Kondo conductance resonances we ask the
reader to refer to reference \cite{MattNT}.

The resonance of ridge ``B'' remains in the superconducting state,
but its conductance is not increased. At first sight this behavior
seems surprising, since resonances indicate a high effective
transmission for which a doubling in conductance is expected in
the unitary limit. This, however, only holds for a symmetrically
coupled junction. Using the zero temperature expression for the
normal state conductance on resonance of a single level
$G_{0}=(2\,e^{2}/h)\;4\Gamma_{L}\Gamma_{R}/(\Gamma_{L}+\Gamma_{R})^2$,
which should hold in the unitary limit, we obtain a relative
asymmetry of the lead coupling of $\Gamma_{L}/\Gamma_{R}=0.37$ (or
the inverse). Between a normal and a superconducting lead the
linear (Andreev) conductance at zero temperature has the form
$G_{0}=(4\,e^{2}/h)\;[2\Gamma_{S}\Gamma_{N}/(\Gamma_{S}^2+\Gamma_{N}^2)]^2$~
\cite{beenakker}. Using the $\Gamma$-ratio determined before one
obtains for the resonance conductance in the superconducting state
$G_{0}=1.69\;e^{2}/h$. This value is only slightly higher than the
one in the normal state and the relative asymmetry of the dot-lead
coupling therefore explains our experimental observation.

Quite remarkably, extending the picture of a non-interacting
resonance in between a normal and a superconductor to finite bias
allows us to determine not only the asymmetry of the two couplings
$\Gamma_{S}$ and $\Gamma_{N}$ but also which one of the two
dominates. In order to do so we follow the BTK-like approach to NS
resonant tunnelling as given by Khlus {\it et al.} \cite{khlus}.
Figure \ref{figure2}(a) shows the results of our simulation where
we took finite temperature into account by setting T=0.1~$\Delta$.
The solid line represents the normal state conductance resonance
for $\Gamma_{L/R}=0.60$ and $\Gamma_{R/L}=0.37 \times 0.60=0.22$
as obtained from the asymmetry and a rough match of the peak
width. The dotted curve represents the resonance in the
superconducting state with a stronger coupling to the
superconductor, i.e. $\Gamma_{S}=0.60$ and $\Gamma_{N}=0.22$. In
this case the simulation data show two pronounced conductance
resonances at $V\approx \pm \Delta/2e$ with a maximum conductance
of approximately $2.7 \:e^{2}/h$. The features are smeared out by
finite temperature, hence resulting in a linear conductance bigger
than the zero-temperature limit $G=1.69 \:e^{2}/h$. The dashed
line shows the second possible scenario in which the normal lead
has a stronger coupling to the quantum dot, thus $\Gamma_{N}=0.60$
and $\Gamma_{S}=0.22$. Here the maximum conductance of order $1.6
\:e^{2}/h$ is reached in the linear-response regime, while
additional structures occur at the position of the gap. Let us now
compare the simulation to the experimental data. Figure
\ref{figure2}(b) shows the measured differential conductance of
ridge ``B'' at 90 mK plotted versus the applied source-drain
voltage for the normal (solid line) and superconducting (dashed
line) case. Indeed the calculated conductance assuming a bigger
coupling to the normal conductor and the measured conductance
agree quite well. Both peak height and the feature at
$V=-\Delta/e$ can be reproduced. The corresponding feature for
positive source-drain voltages, though, is washed out by the
asymmetric shape of the measured Kondo resonance. Yet we are able
to conclude that for our sample the normal lead exhibits a better
coupling to the dot than the superconducting lead with an
asymmetry of $\Gamma_{N}\approx \Gamma_{S}/0.37$.

In this article we studied a carbon nanotube quantum dot in the
Kondo regime coupled to a normal and a superconductor. In the case
of $T_{K}<\Delta$ the Kondo ridge at zero bias disappears and
peaks at the position of the gap occur. In the case $T_{K}>\Delta$
the Kondo resonance persists but does not show an enhancement of
the conductance compared to the normal state which we attribute to
an asymmetric coupling of the electrodes. Agreement was found when
comparing the data to a simple model of a resonant NS-interface.
Future experiments will have to (a) clarify whether the Kondo
resonance can actually be enhanced in presence of the
superconducting electrode by tuning the coupling asymmetry
$\Gamma_{S}/\Gamma_{N}$ and (b) explore the possibility of
generating pairs of entangled electrons by making use of nanotubes
coupled to normal and superconducting leads~\cite{NTentangler}.

\section*{Acknowledgments}
We thank M. Buitelaar, B. Choi, L. Grueter and S. Sahoo for
experimental help and A. Clerk, J. Cuevas, A. Levy Yeyati and P.
Recher for discussions. We thank L. Forr\'o for the MWNT material
and J. Gobrecht for the oxidized Si substrates. This work has been
supported by the Swiss NFS and the NCCR on Nanoscience.

\end{document}